\documentclass[useAMS,usenatbib]{mn2e}

\usepackage[fleqn]{amsmath}
\usepackage{booktabs}
\usepackage{natbib}
\usepackage{graphicx}
\usepackage{listings}
\usepackage{bookmark}
\usepackage{hyperref}
\hypersetup{colorlinks=true, urlcolor=blue, citecolor=blue}
\title[Estimating the spiral pattern speed in the Milky Way]{A new method for estimating the pattern speed of spiral structure in the Milky Way}

\author[T. C. Junqueira et al.]{T. C. Junqueira$^{1}$\thanks{E-mail: tjunqueira@aip.de}, C.  Chiappini$^{1}$, J. R. D. L\'epine$^{2}$, I. Minchev$^{1}$, B. X. Santiago$^{3,4}$
\\
\\
$^{1}$Leibniz-Institut f\"ur Astrophysik Potsdam, An der Sternwarte 16, 14482 Potsdam, Germany
\\
$^{2}$Instituto de Astronomia, Geof\'isica e Ci\^encias Atmosf\'ericas, Universidade de S\~ao Paulo, Cidade Universit\'aria, \\
S\~ao Paulo 05508-090, SP, Brazil
\\
$^{3}$Instituto de F\'isica, UFRGS, CP 15051, Porto Alegre, RS 91501\-970, Brazil 
\\
$^{4}$Laborat\'orio Interinstitucional de e-Astronomia-LIneA, Rua Gal. Jos\'e Cristino 77, Rio de Janeiro, RJ 20921-400, Brazil} 

\begin{document}

\date{}


\maketitle

\label{firstpage}

\begin{abstract}
In the last few decades many efforts have been made to understand the effect of spiral arms on the gas and stellar dynamics in the Milky Way disc. One of the fundamental parameters of the spiral structure is its angular velocity, 
or pattern speed $\Omega_p$, which determines the location of resonances in the disc and the spirals' radial extent. The most direct method for estimating the pattern speed relies on backward integration techniques, trying to 
locate the stellar birthplace of open clusters. Here we propose a new method based on the interaction between the spiral arms and the stars in the disc. Using a sample of  around 500 open clusters from the 
{\it New Catalogue of Optically Visible Open Clusters and Candidates}, and a sample of 500 giant stars observed by APOGEE,  we find $\Omega_p = 23.0\pm0.5$ km s$^{-1}$ kpc$^{-1}$, for a local standard of rest rotation  
$V_0=220$~km s$^{-1}$ and solar radius $R_0=8.0$~kpc. Exploring a range in $V_0$ and $R_0$ within the acceptable values, 200-240 km s$^{-1}$ and 7.5-8.5 kpc, respectively, results only in a small change in our estimate of
$\Omega_p$, that is within the error. Our result is in close agreement with a 
number of studies which suggest values in the range 20-25 km s$^{-1}$ kpc$^{-1}$. An advantage of our method is that we do not need knowledge of the stellar age, unlike in the case of the birthplace method, which allows us to use 
data from large Galactic surveys. The precision of our method will be improved once larger samples of disk stars with spectroscopic information will become available thanks to future surveys such as 4MOST. 
\end{abstract}


\section{INTRODUCTION}

The Milky Way (MW) has long been known to posses spiral structure, but its fundamental nature is still under debate today. In the classical spiral structure theory of \citet{Lin_Shu1964}, as well as other models which consider that 
the spiral arms are caused by the crowding of stellar orbits \citep[e.g.][]{ Kalnajs1973,Contopoulos1986,Pichardo2003,Junqueira2013}, the spiral arm pattern rotates like a rigid body with a well defined angular velocity 
$\Omega_p$. In these models $\Omega_p$ is usually treated as a free parameter to be determined by observations. However, its value has a crucial importance on the understanding of Galactic dynamics and evolution, since it 
determines the place of resonances in the disc, for a given rotation curve. In addition to this challenge, there are also theories of spiral arms claiming that the pattern speed and pitch angle are variable \citep{Toomre1981}, 
that spirals are transient phenomena on a rotational time scale (e.g., \citealt{Sellwood2002}), or that there exist at any time several spiral sets with distinct pattern speeds overlapping in radius e.g., 
\citealt{Masset1997,Merrifield2006,Quillen2011,Minchev2012}, or even that the spiral arms are stochastic phenomena e.g., \citealt{Patsis2006}.

Looking at the existing theories, we can see that it is not clear whether the $\Omega_p$ can be described by a multiple, transient or a constant pattern speed. However, some authors have shown that the corotation radius 
is close to the solar orbit \citep{Marochnik1972,Creze1973,Mishurov1999,Dias2005}. \cite{Amores2009} associated a gap in the Galactic HI distribution close to the corotation radius, while \cite{Scarano2013} 
showed a break in the radial metallicity gradient close to the corotation radius for many external galaxies. This suggests that we have dominant spiral arms with a constant pattern speed for, at least, a few billion years, 
which do not support models with transient spiral arms that survive for only a few galactic rotations at the solar radius. Thus, the approximation of a single and steady pattern speed is a useful first step to see whether this assumption is consistent 
with available data.

The most direct method to measure the pattern speed of the MW relies on the birthplaces of the observed open clusters. It is done by integrating backwards in time their orbits according to their known location, ages and circular 
velocity in the disk. Assuming that the open clusters are born in spiral arms, the distribution of birthplace for some age bins should be spiral-like (where we assume logarithmic spirals), and by comparing 
the spiral patterns obtained from different ages bins, the rotation rate of the spiral arms can be estimated. This is a valuable way to measure the pattern speed, but determining ages for a large open cluster sample is not an easy 
task and may introduce large errors.

In the present work, we introduce a new method to determine the angular velocity $\Omega_p$ based on the interaction between the stars and the spiral arms, that cause an exchanged of energy and angular momentum (we use 
energy and angular momentum per mass unit, but for simplicity we keep the terminology energy and angular momentum). The only assumption made in this method is that the initial energy is the energy of a circular orbit 
placed at the mean radius $r_m$. We do not give any information a priori about the spiral arms, which is already included in the velocities components of each object. Moreover, we do not need to know the ages or make 
any assumption about the spiral arms shape. In addition, the fact that we do not have to make use of ages allows us to use observational data that provide only positions and kinematics information, and hence can be a powerful 
method in the era of large spectroscopic surveys such APOGEE-2 (as part of SDSS-IV - \cite{Sobeck2014}), and future very-large ones as 4MOST\footnote{4-m multi-object spectrograph telescope - \cite{Jong2012}} and 
WEAVE\footnote{http://www.ing.iac.es/weave/index.html}.

The organization of this paper is as follows: in Section 2 we present the new method for measuring the spiral pattern speed of the Galaxy. In this same Section we provide the rotation curve used to compute the Galactic potential and how the 
mean radius is calculated for each object, as well as some tests made to validate the method. In Section 3, we describe the data used to compute $\Omega_p$. In Section 4, we compare our result with the ones found in the literature and 
discuss possible sources of errors. Concluding remarks can be found in Section 5.

\section{METHOD}
\label{mtd}

As a star travels around the Galactic center it interacts with the spiral arms, which results in exchange of angular momentum and energy. The rate at which this happens depends on the relative angular velocity 
$\Omega - \Omega_p$, where $\Omega$ is the observed stellar angular velocity. Thus, in an inertial frame of reference, the energy of a star varies due to perturbations caused by the presence of the spiral arms but, if there is 
only one pattern speed, we can find a Hamiltonian that is time independent. This Hamiltonian system lies on the frame of reference of the spiral arms, which is a non-inertial frame of reference. This is known as the Jacobian 
of the system \citep{Binney2008}, written as:

\begin{equation}
\label{ha1} 
 H = E - \Omega_pJ, 
\end{equation}

\noindent where $H$ is the energy in the non-inertial frame of reference, while $E$ and $J$ are the energy and the angular momentum in the inertial frame of reference, respectively. The energy of star $E$ is given by:

\begin{equation}
\label{En} 
E = \frac{1}{2}\left(U^2 + \frac{J^2}{r^2}\right) + \Phi_0(r) + \Phi_1(r,\theta),
\end{equation}

\noindent and

\begin{equation}
\label{Jn} 
J = r^2\Omega,
\end{equation}

\noindent $U$ is the radial velocity toward the Galactic center, $r$ is the star's distance from the Galactic center, $\Omega = V/r$ where $V$ is the rotational velocity, 
$\Phi_0(r)$ is the axisymmetric potential and $\Phi_1(r,\theta)$ is the perturbative potential due to spiral arms. Once the total energy $H$ is 
conserved ($\Delta H = 0$) we can derive, from Eq.~\ref{ha1}, that the energy variation for each star is proportional to the angular momentum variation;

\begin{equation}
\label{ha2} 
 \Delta E =  \Omega_p\Delta J.
\end{equation}

Eq. \ref{ha2} is not new and it was obtained also by \cite{Bell1972} and \cite{Sellwood2002}. The process of steady angular momentum transfer between the spiral density and a star in resonant motion with the perturbation 
is understood in the following way: the loss (gain) of angular momentum $\Delta J$ by a star at the inner Lindblad resonance is accompanied by the loss of energy $\Omega_{p}\Delta J$; at the same time, the 
change in orbital energy, relative to the circular motion, is $\Omega\Delta J$, which is less than $\Omega_{p}\Delta J$. The connection between the orbital energy $E$ and angular momentum $J$ 
is shown by Eq.~\ref{En}, where for a pure circular motion it becomes $E_c = J^2/(2r^2) + \Phi_0(r)$, thus any variation in orbital energy leads to a variation in angular momentum at a rate proportional to the 
stellar angular velocity $\Omega$. Thus the amount of energy in radial direction $\Delta E_{r}$ acquired by the star appears as non-circular motion:

\begin{equation}
\label{er}
\Delta E_{r}=(\Omega_{p}-\Omega)\Delta J,
\end{equation}


We can merge Eqs.~\ref{ha2} and~\ref{er} into the following:

\begin{equation}
\label{op}
\Delta E_{r} + \Delta E + \Omega\Delta J = 2\Delta J\Omega_{p}.
\end{equation}

The challenge here is to find $\Delta E_{r}$, $\Delta E$, and $\Delta J$ for a real set of stars in the Galaxy. The first consideration that we can make is that the axisymmetrical potential and the kinetic energy are larger than 
the perturbation, thus we might ignore the term $\Phi_1(r,\theta)$ from Eq.~\ref{En}. This is justifiable because the residual velocities of the objects in comparison with the pure circular ones already carries information about 
the perturbing field, and another important reason is that we are not giving any prior information about the spiral arms. Thus we can rewrite Eq.~\ref{En} ignoring the perturbative term;

\begin{equation}
\label{En2} 
E \simeq \frac{1}{2}\left(U^2 + \frac{J^2}{r^2}\right) + \Phi_0(r).
\end{equation}

When we observe a star we just have access to its energy and angular momentum at one specific period, in other words, we cannot track a star for a few millions of years to see how it will change its energy and angular momentum over time. 
In an idealized case, where we know exactly the values of $\Delta E$ and $\Delta J$ (e.g. in a simulation), we could use Eq.~\ref{ha2} to recover $\Omega_p$. However, the problem of not knowing the exact values of $\Delta E$ and 
$\Delta J$ introduces errors in this computation. This is why we make use of Eq.~\ref{op}, which is the combination of Eqs.~\ref{ha2} and~\ref{er}, forcing both equations to be satisfied giving us a better result. 
We also average the values of $y=\Delta E_{r} + \Delta E + \Omega\Delta J$ and $x = 2\Delta J$ for all stars inside a bin,  which improves our results as explained below. It is important
to emphasize that we bin the mean radius of the stars and not the actual radial position.    

To solve the problem of the unknown $\Delta E$ and $\Delta J$, we assume that the initial energy $E_0$ and angular momentum $J_0$ of each star is the energy of a circular orbit at the mean radius $r{_m}$ (Eq.~\ref{rm} gives the 
definition of $r_m$):

\begin{equation}
\label{Ec} 
E_0 = \frac{J_{0}(r_{m})^2}{2r_{m}^2} + \Phi_0(r_{m}),
\end{equation}

\noindent with

\begin{equation}
\label{J0} 
J_0 = r_mV_c(r_m),
\end{equation}

\noindent where $V_c$ is the rotation curve supplied by Eq.~\ref{eq:v_L2011}. Now we have all the necessary ingredients to compute the variation in energy and angular momentum:

\begin{equation}
\label{dele} 
\Delta E_i =  E_i - E_{0_i} = \frac{U_{i}^2}{2} + \frac{V_{i}^2-V_{c}^2(r_{m_{i}})}{2} + \Delta\Phi_0(r_i-r_{m_i}),
\end{equation}

\begin{equation}
\label{delj} 
\Delta J_i =  J_i - J_{0_i} = r_iV_i - r_{m_i}V_c(r_{m_i}),
\end{equation}

\noindent and

\begin{equation}
\label{deler} 
\Delta E_{r_i} =  \frac{U_i^2 - U_{max}^2}{2}.
\end{equation}

The sub-index $i$ refers to each star, $V_i$ and $U_i$ are respectively, the observed circular velocity and the radial velocity with respect to the center of the Galaxy, and $U_{max}$ is the maximum radial velocity, 
that happens when the star crosses $r_m$. Both velocities $V_i$ and $U_i$, are not directly observables, what we observe directly are the proper motion and the heliocentric velocity, to make this transformation we follow 
\cite{Johnson1987}. To correct the motion due to the local standard rest (LSR) we used the values from \cite{Schonrich2010}; $u_0 = -11.1$ km s$^{-1}$ and $v_0 = 12.24$ km s$^{-1}$.

The next step is to bin the mean radius in intervals of $\Delta r = 0.2$ kpc from 5 up to 14 kpc. The reason for it is that stars with similar mean radius are expected to have almost 
the same energy $H$, which reflect the initial energy of a circular orbit in which the star could have come from (see Section \ref{Am} for a better explanation). Therefore, inside each bin we have a number $N$ of stars within $r < r_m < r + \Delta r$ and we 
average $\Delta E_i$, $\Delta J_i$ and $\Delta E_{r_i}$ for all the $N$ stars within the bin. Thus Eq.~\ref{op} can be rewritten as:

\begin{equation}
\label{op2}
\Delta \overline{E} = 2\Omega_{p}\Delta\overline{J},
\end{equation}

\noindent where $\Delta\overline{E}$ and $\Delta\overline{J}$ give us the energies and the angular momentum variation in each bin;

\begin{equation}
\label{delem} 
\Delta \overline{E} =  \frac{1}{N}\sum_{i=1}^{N}(\Delta E_i + \Delta E_{r_i} + \Omega_i\Delta J_i),
\end{equation}

\noindent and

\begin{equation}
\label{deljm} 
\Delta \overline{J} =  \frac{1}{N}\sum_{i=1}^{N}\Delta J_i.
\end{equation}

Thereby, using Eq.~\ref{op2}, we can recover the value of $\Omega_p$ without giving any previous input about the perturbation by fitting a first degree equation $y = ax + b$. Where, $y = \Delta \overline{E}$ and 
$x = 2\Delta\overline{J}$, with $a$ and $b$ as free parameters. The slope of this fit give us the value of $\Omega_p$.

In Section 2.1 we show how we compute the mean radius and the rotation curve. In Section 2.2 we carry out tests based on simulated particles in order to show our method works, and to estimate the expected uncertainty on the 
retrieved $\Omega_p$ parameter.

\begin{table*}
 \centering
  \caption{Parameters of the rotation curve (Eqs.~\ref{eq:v_L2011} and~\ref{eq:funct_mrc}).}
  \label{tab:param_rotcurv}
  \begin{tabular}{@{}ccccccccc@{}}
  \hline
  \hline
   
  \multicolumn{1}{c}{$\balpha$} &
	\multicolumn{1}{c}{$\bbeta$} &
	\multicolumn{1}{c}{$\bgamma$} &
	\multicolumn{1}{c}{$\bdelta$} &
	\multicolumn{1}{c}{$\bepsilon$} &
	\multicolumn{1}{c}{$\boldeta$} &
	\multicolumn{1}{c}{$\mathbf{A_{\mathrm{mrc}}}$} & 
	\multicolumn{1}{c}{$\mathbf{R_{\mathrm{mrc}}}$} &
	\multicolumn{1}{c}{$\bsigma_{\mathrm{mrc}}$} \\
	
	\multicolumn{1}{c}{[km s$^{-1}$]} &
	\multicolumn{1}{c}{[kpc]} &
	\multicolumn{1}{c}{[kpc]} &
	\multicolumn{1}{c}{[km s$^{-1}$]} &
	\multicolumn{1}{c}{[kpc]} &
	\multicolumn{1}{c}{[kpc]} &
	\multicolumn{1}{c}{[km s$^{-1}$]} &
	\multicolumn{1}{c}{[kpc]} &
	\multicolumn{1}{c}{[kpc]} \\
 \hline
 250 &120 &3.4 &360 &3.1 &0.09 &20 &8.8 &0.8 \\
\hline
\end{tabular}
\end{table*} 

\subsection{Assumptions of the method and uncertainties}
\label{Am}

We critically discuss the main assumptions of our method and their impact on the recovered $\Omega_p$ value. The main assumption of our method is that there is a dominant pattern speed that lasts, at least, for a few billion 
years. Notice that all existing methods are based on this premise.

As stated before, in a real sample we don't have access to the true variation in energy and angular momentum of each star. To overcome this problem another main assumption made here is that the initial energy $E_0$ and the initial 
angular momentum $J_0$, come from a circular orbit placed at the mean radius (see Fig.~\ref{fig1}). In a pure circular orbit all the stars at the same radius have the same energy and angular momentum as those quantities depend 
only on the stellar position (see Eqs.~\ref{Ec} and \ref{J0}). We then split our sample in bins of mean radius (as stars with similar mean radius will have, according to our assumption, similar initial energy $E_0$ and angular 
momentum $J_0$). The bin width we chose was 0.2 kpc, to assure that the Eqs.~\ref{Ec} and \ref{J0} do not vary too much and at the same time we could have enough stars (at least more than 5) in each bin. We next discuss the 
impact radial migration would have on this important assumption of our method. 

\cite{Minchev2013} showed that in a simulated disk similar to the Milky Way migration is a global process, significantly affecting the entire disk. How would this affect our results, in particular the relation 
$\Omega_p = \Delta E/\Delta J$? We expect that the impact of migration on our method will be small for the following reasons. \cite{Minchev2012} showed that, due to conservation of vertical and radial actions, migrators arrive 
at a new radial location with orbital properties very similar to the stars which did not migrate. Therefore, the $\Delta E$ and $ \Delta J$ values at the final migration time are expected to be very similar to those of the 
local non-migrators. An exception to this rule would come from stars  currently in the process of migration. While those could be a large number at high redshift due to the strong effect of external perturbers 
(e.g., large infalling satellites), it should be expected that at present these ''migrators in action'' do not constitute a significant fraction of the stars found in a given radial bin. 

Another possible source of error can result from kinematically hot stars, i.e., stars with high eccentricity, for which our main assumption that the initial energy $E_0$ and the initial angular 
momentum $J_0$, come from a circular orbit at the mean radius $r_m$, starts being imprecise. This is the main reason for using averaged $\Delta E$ and $\Delta J$ values for stars sharing the same bin in mean radius. 
Because the stars with hot kinematics are mostly old, we adopt a) a sample of open clusters for which only $8\%$ have ages above 1 Gyr (see Fig.~\ref{figX3}), b) a stellar sample confined to the 
Galactic plane, mostly dominated by thin disk stars and c) spiral arms have the strongest dynamical effect in the disk midplane and thus stars with low vertical oscillations are the best tracers for the arms. 
As we discuss in the Results Section, the fact that we obtain very similar results from both samples, show the impact of the above mentioned shortcomings (radial migration and stars on eccentric orbits) in our method to 
be minor.

Finally, the main source of error in our method comes from the computation of $\Delta E$ and $\Delta J$. To illustrate how the errors from $\Delta E$ and $\Delta J$ affect the measurement of $\Omega_p$, we propagate the 
errors using the Eq.~\ref{ha2} and we derived the equation bellow: 

\begin{equation}
\label{error}
\sigma_{\Omega_p}=\frac{1}{\Delta J}\sqrt{\sigma_{\Delta E}^2 + \frac{\Delta E^2}{\Delta J^2}\sigma_{\Delta J}^2}.
\end{equation}

It simplify the analysis once we assume that the major errors come only from $\Delta E$ and $\Delta J$. Here $\sigma_{\Delta J}$ and $\sigma_{\Delta E}$ are the errors from these two variables. Now let's assume that the errors follow 
the relation; $\sigma_{\Delta E}=\alpha\Delta E$ and $\sigma_{\Delta J}=\beta\Delta J$.  This tell us that the errors are proportional to the own variation of energy and angular momentum, respectively. This makes sense because $E_0$ 
and $J_0$ become less precise when the variation of $\Delta E$ and $\Delta J$ are larger. Thus, for the errors following the given definition we can rewrite the equation above as:

\begin{equation}
\label{error2}
\sigma_{\Omega_p}=\frac{\Delta E}{\Delta J}\sqrt{\alpha^2 + \beta^2} = \Omega_p\sqrt{\alpha^2 + \beta^2}.
\end{equation} 

For a single star the parameters, $\alpha$ and $\beta$, can be larger than one, which leads to an error greater than $100\%$ in $\Omega_p$. However, the errors in $\alpha$ and $\beta$ are much smaller once we use a large number 
of stars averaged on a particular mean radius bin. Indeed, for the test sample $\alpha$ and $\beta$ are $\approx$ 0.1, which can be seeing in the errors bar for $\Delta E$ in Fig.~\ref{fig2}, for $\Delta J$ the magnitude of the errors bar are the same. Thus, from Eq.~\ref{error2} with 
$\alpha$ = $\beta$ =  0.1 and $\Omega_p$ = 23 km s$^{-1}$ kpc$^{-1}$ we have an error of $\sigma_{\Omega_p}$ = 3 km s$^{-1}$ kpc$^{-1}$. In a more general way $\sigma_{\Omega_p} = 0.14\Omega_p$ km s$^{-1}$ kpc$^{-1}$

In summary, the oldest the population, the less precise is our method. On the other hand, a very young population ($<$10 Myrs), which most probably would not have had enough time to interact with the spiral arms, because of 
the transformation of dissipational to collisionless dynamics, can also lead to uncertain results. An optimal sample would be composed of stars with ages between 50 Myrs and a few Gyrs. Finally, as discussed before we expect larger errors for larger values of $\Delta E$ and $\Delta J$. Hence, larger uncertainties 
should be expected for stars migrating from the resonances (i.e. inner or outer Lindblad resonance, ILR or OLR), as these stars might have increased significantly their energy and and angular momentum. The two samples adopted here 
were chosen with the aim to minimize these effects (see Section \ref{data}). Finally, we notice that a  large number of stars per mean radius bin improves our determination of $\Omega_p$.

\subsection{Mean radius and the rotation curve}

In order to find the mean radius we adopt a model for the axisymmetric galactic potential that reproduces the general behavior of the rotation curve of the Galaxy. We use an analytical expression to represent the circular 
velocity as a function of Galactic radius, conveniently fitted by exponential in the form (units are km s$^{-1}$ and kpc):

\begin{equation}
\label{eq:v_L2011}
V_{c}(r)=\alpha\exp\left[-\frac{r}{\beta}-\left(\frac{\gamma}{r}\right)^{2}\right]+\delta\exp\left[-\frac{r}{\epsilon}-\frac{\eta}{r}\right]+f_{\mathrm{mrc}}(r),
\end{equation}

\noindent with

\begin{equation}
f_{\mathrm{mrc}}(r)=-A_{\mathrm{mrc}}\exp\left[-\frac{1}{2}\left(\frac{r-R_{\mathrm{mrc}}}{\sigma_{\mathrm{mrc}}}\right)^{2}\right],
\label{eq:funct_mrc}
\end{equation}

\noindent where $A_{\mathrm{mrc}}$ is the amplitude and $\sigma_{\mathrm{mrc}}$ is the half-width of the minimum centered at the radius $R_{\mathrm{mrc}}$. We verified that the adopted depth of the 
minimum in Eq.~\ref{eq:funct_mrc}  does not have any measurable effect on the value of $\Omega_p$ obtained  in this work. The rotation curve given by the expressions in Eqs.~\ref{eq:v_L2011} and~\ref{eq:funct_mrc} is close to that 
derived by \citet{Fich1989} and is also similar to the ones previously used by, e.g., \citealt{Lepine2008}; ALM; \citealt{Lepine2011a}. The interpretation of a similar curve in terms of components of the Galaxy is given 
by \citet{Lepine2000}. Table~\ref{tab:param_rotcurv} gives the values of the parameters chosen to reproduce the rotation curve of the Milky Way. For the Galactocentric distance of the Sun, we adopt $R_{0}$ = 8.0 kpc. 
The circular velocity at $R_{0}$ resultant from Eq.~\ref{eq:v_L2011} is $V_{0}=210$ km s$^{-1}$, for a peculiar velocity of the Sun in the direction of Galactic rotation v$_0 = 12.24$ km s$^{-1}$ the velocity with respect to 
the local standard rest is $V_{LSR}\simeq220$ km s$^{-1}$. For more details and also a theoretical description about the minimum close to the solar position see \cite{Barros2013}. 

As we restrict our study to orbits in the galactic plane, the axisymmetric potential can be derived directly from the rotation curve:

\begin{equation}
\label{eq:Phiax}
\Phi_{0}(r)=\int{\frac{V^{2}_{c}}{r}dr}.
\end{equation}

\begin{figure}
\includegraphics[scale=.5,angle=0]{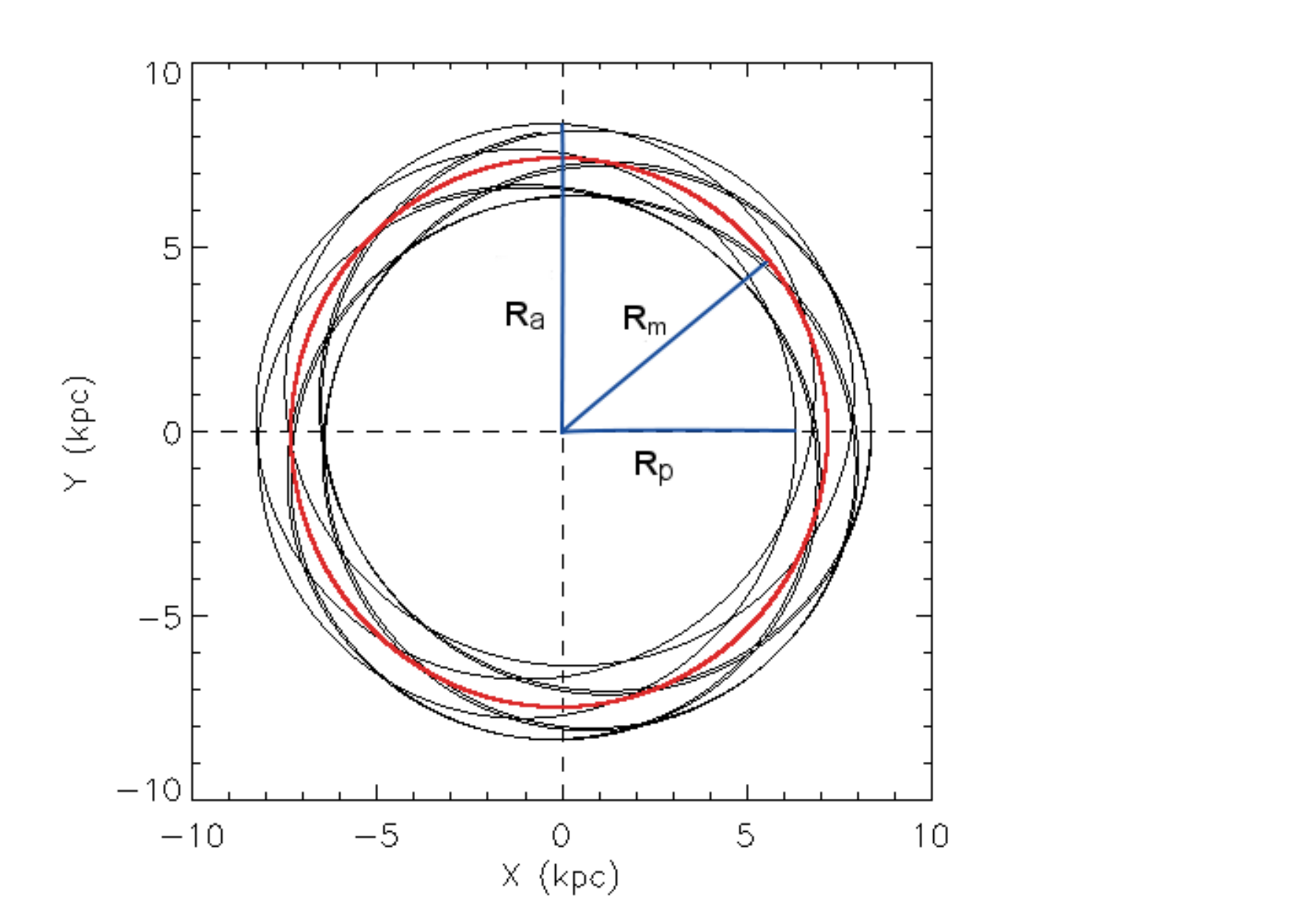}
\caption{This figure illustrates a typical orbit, in an inertial frame of reference, where the star follows elliptical orbits which are confined in a circular region limited by the pericenter ($r_p$) and the 
apocenter ($r_a$). The black lines show the stellar orbit, the red line is the position of the mean radius $r_m$.}
\label{fig1}
\end{figure}

We integrated the stellar orbits for 2 Gyr under the influence of the potential $\Phi_0(r)$, excluding any perturbations. This allowed us to find the apocenter ($r_a$) and pericenter ($r_p$) radius. 
The mean radius is then given by;

\begin{equation}
\label{rm}
r_m=\frac{r_a+r_p}{2}.
\end{equation}

Fig. \ref{fig1} shows an example of an orbit with the apocenter, pericenter and mean radius of a star in the Galactic potential.

\subsection{Application of the method to modeled data}
\label{amd}

In order to test our method we integrate the orbits of 500 test-particles for 2 Gyr under the influence of a perturbing potential with a given value of $\Omega_p$. We chose 500 particles to match with the number of stars we have available 
in each sample, approximately. With future data, the increase of these number and a better distributed in the Galactic plane, could improve the results, as we will discuss later in the Sec.~\ref{data}. The potential and parameters that we used for the spiral arms are described in \cite{Junqueira2013} (their Eq. 6 with the parameter values given on their Tab. 1). The axisymmetric potential $\Phi_0$ comes from Eq.~\ref{eq:Phiax}, with the rotation curve $V_c$ given by 
Eq.~\ref{eq:v_L2011}, where we set $\Phi_0(100)$ = 0 at $r$ = 100 kpc to find the constant of integration. Initially, the test-particles were distributed between 6 and 12 kpc with random azimuthal positions, and 
initial circular velocities corresponding to the rotation curve $V_c$. The test-particles have initial radial velocities $U$, given by a Gaussian shape in each bin of radius and the half-width is the velocity dispersion $\sigma_r$ 
that follows a radial profile given by Eq.~\ref{sigr}. 

\begin{figure}
\includegraphics[scale=.6,angle=0]{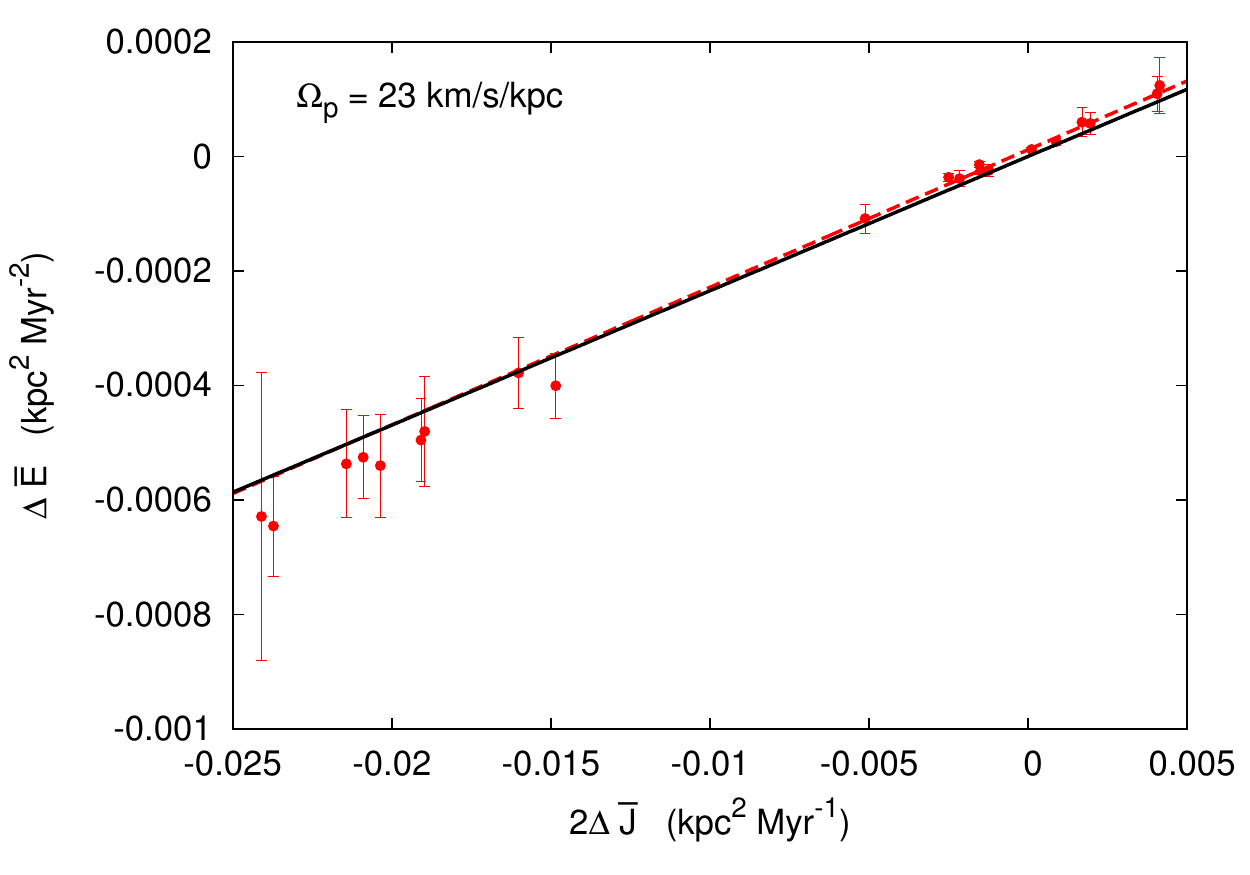}
\includegraphics[scale=.6,angle=0]{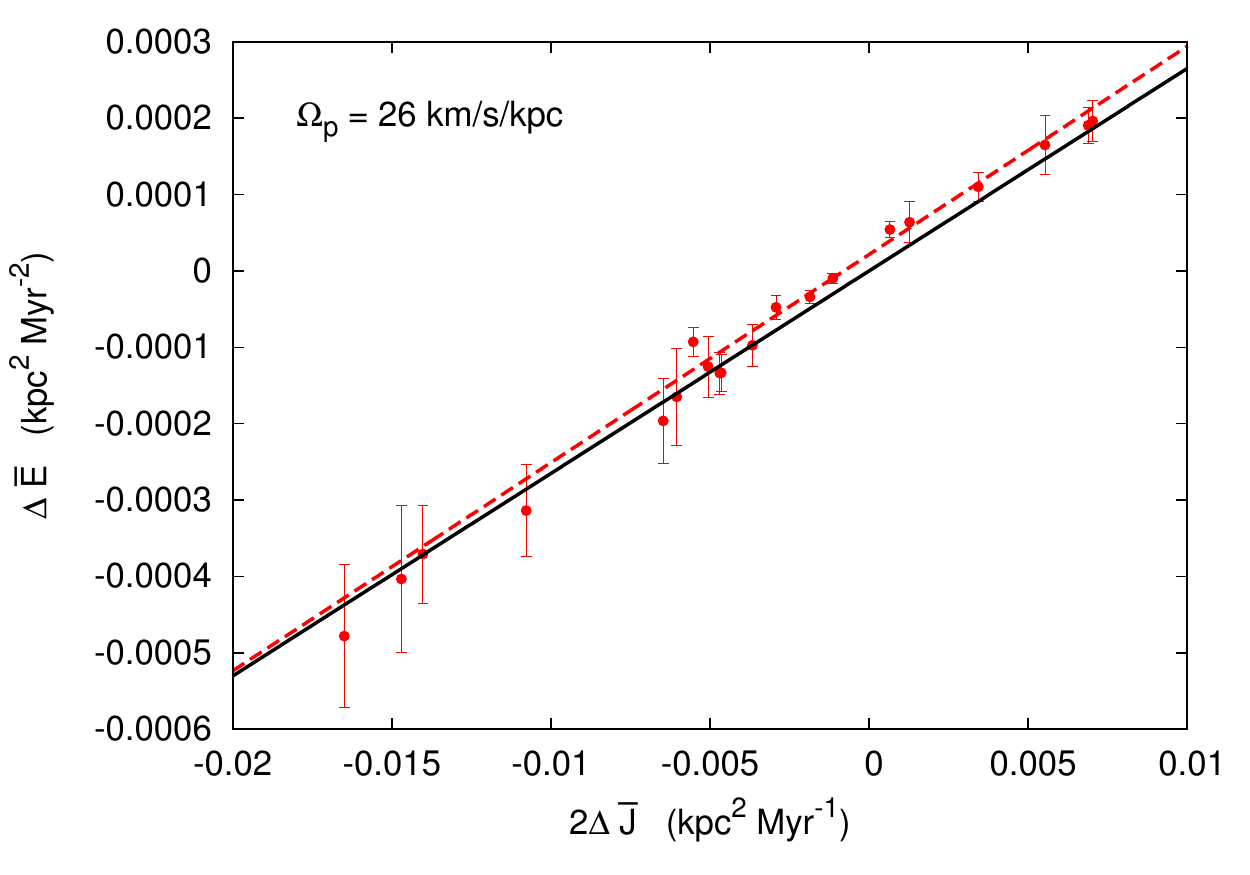}
\includegraphics[scale=.6,angle=0]{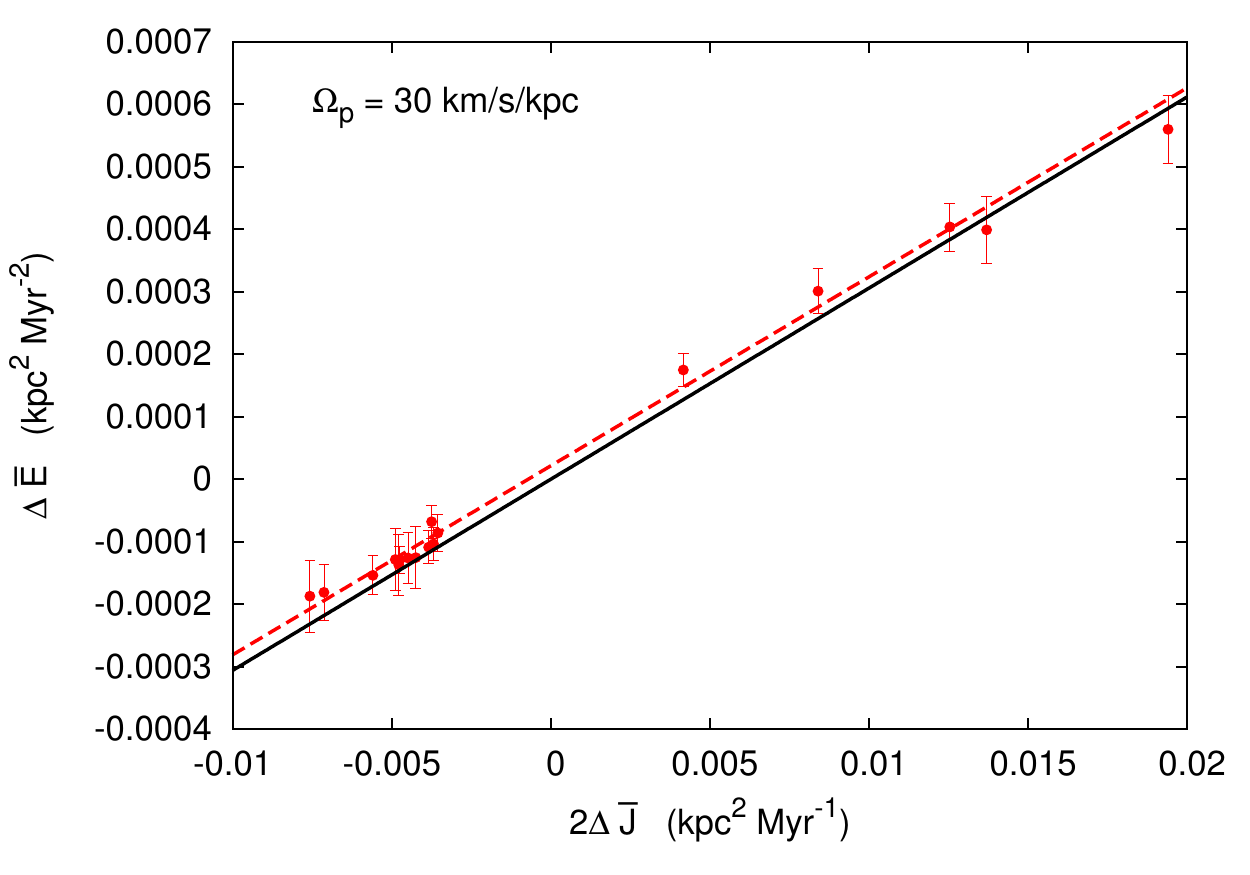}
\caption{This figure shows the linear correlation between $2\Delta \overline{J}$ and $\Delta \overline{E}$ for three different values of $\Omega_p$. The red dots are the values computed from Eqs.~\ref{delem} and 
\ref{deljm}, as explained in Sec.~\ref{mtd}, the dashed red lines are the least-squares fit on these points, where the slope give us the value of $\Omega_p$ (shown on Tab.~\ref{tb2}), and the black lines are the linear relation for 
the original value of $\Omega_p$.}
\label{fig2}
\end{figure}

\begin{equation}
\label{sigr}
\sigma_r=\sigma_{r_0}\exp\left(\frac{r_0-r}{5}\right),
\end{equation}

\noindent with $r_0 = 8$ kpc at the solar radius and a velocity dispersion $\sigma_{r_0} = 6.5$ km/s. These values are compatible with 
the amplitudes of the perturbation velocities due to the spiral waves found in the literature (e.g. \citealt{Burton1971}, \citealt{Mishurov1997}, \citealt{Bobylev2010}). 
They are also similar to the velocity dispersion of the youngest Hipparcos stars (\citealt{Aumer2009}).

\begin{table}
\caption{The first column shows the input value for $\Omega_p$ and the second column shows the value of $\Omega_p$ recovered by our method.
All values in this table have units of km s$^{-1}$ kpc$ ^{-1}$.}
\label{tb2}
\begin{center}
\begin{tabular}{|c|c|}
\toprule
Input $\Omega_p$ & Recovered $\Omega_p$ \\
\hline\hline
23 & 23.5$\pm$0.9 \\
26 & 26.7$\pm$1.0 \\
30 & 29.7$\pm$0.7 \\
\bottomrule
\end{tabular}
\end{center}
\end{table}

The second step is to recover the value of $\Omega_p$ from this synthetic sample that we generated. To do that we selected the particles from different snapshots in order to simulate different ages, as 
we have in a real sample. After that, we re-integrate the orbits using only the axisymmetric potential to find the mean radius of each particle. Then we use the method explained before to compute the 
Eqs.~\ref{delem} and~\ref{deljm}. In Fig.~\ref{fig2} we show the results obtained for different values of $\Omega_p$, where the error bars are the RMS (root mean square) of each bin. Table~\ref{tb2} summarizes the results; the first 
column is the input values of $\Omega_p$, the second column shows the recovered values, that are given by the slope of a least-squares fit shown in Fig.~\ref{fig2}, with the respective asymptotic standard errors.
We can see that our method is able to recover the value of $\Omega_p$ down to around 4\% precision. 

In Sec.~\ref{res} we will apply this method to a sample of open clusters and stars from Apogee catalog to compute the $\Omega_p$ of the MW spiral arms.

\section{THE DATA}
\label{data}

In this Section we describe the two data sets we adopt to illustrate our new proposed method for deriving the pattern speed of the spiral structure of the Milky Way. As discussed in Section~\ref{amd} we are interested in a sample 
of stars and/or clusters dominated by young-intermediate age objects, mostly close to the Galactic plane and for which we have have good distances, radial velocities and proper motion information. Here we adopt two samples, which 
have complementary advantages and disadvantages. In this way we are able to illustrate the robustness of our method. Indeed, as we will see in the Results section, the two samples lead to essentially the same results, despite their different 
azimuthal and age coverage. We now describe each of the samples.     

\subsection{Open Clusters}

Open clusters play an important role on the study of Galactic dynamics, because they are mostly concentrated in the disc plane. Thus, we can use them to find evidences about the kinematic and 
evolution of the MW's disc. 

\begin{figure}
\includegraphics[scale=.6,angle=0]{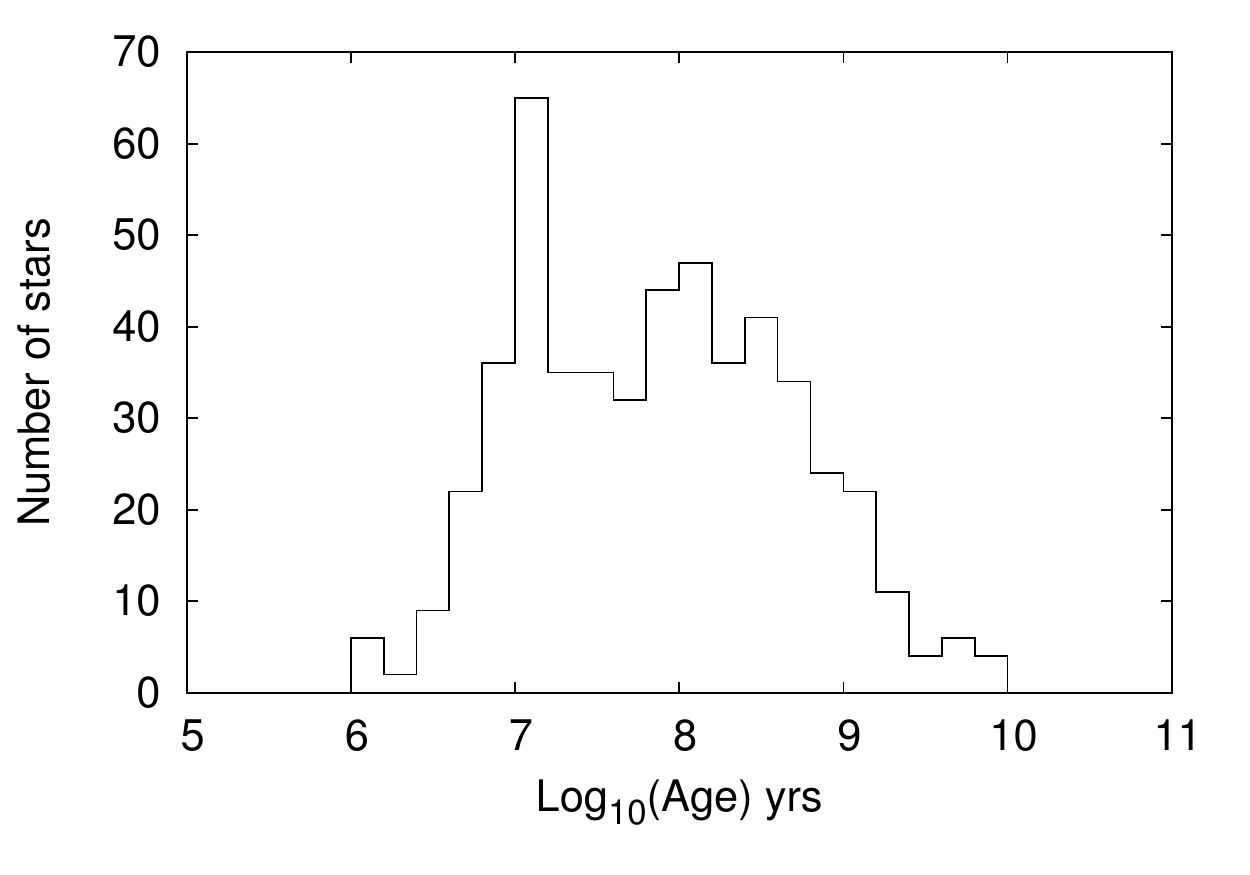}
\caption{Logarithmic distribution of the ages for the OC's sample.}
\label{figX3}
\end{figure}

The open clusters that we used to measure the pattern speed of the spiral arms belong to the ``New Catalog of Optically Visible Open Clusters and Candidates", published by 
\cite{Dias2002}(DAML02)\footnote{\url{http://www.astro.iag.usp.br/ocdb/}}, version 3.3. This catalog is an update from the previous ones by, \cite{Lynga1987} and \cite{Mermilliod1995}, and contains 2140 objects with measured 
parameters such as distance (for 74.5\% of the sample), age (74.5\%), proper motion (54.7\%) and radial velocity (24.2\%). In this work we used 513 open clusters from this catalog, which have distance, proper motion and radial 
velocity available simultaneously. Fig.~\ref{figX3} shows the age distribution of our OC sample. It can be seen that most of the objects are confined in the 10-1000 Myr age range which is an ideal age range for applying 
our method (see discussion in Section \ref{amd}). Finally, the mean radius distribution of our OC sample is shown in Fig.~\ref{figX} (solid black line). The OC sample is concentrated in the 7-10 kpc mean radius range where the 
percentage of young stars coming from the resonance regions is expected to be small \citep[see][]{Minchev2013,Minchev2014} .For the OCs, \cite{Paunzen2006} show a limit of 
20$\%$ in the errors for distances, which are similar to APOGEE HQ sample, and the errors in proper motion for $90\%$ of our OCs sample are less than 1 mas/yr. The uncertainties in radial velocities 
are less than $5\%$.

\subsection{APOGEE}

The Apache Point Observatory Galactic Evolution Experiment (APOGEE; \citealt{Allende2008}; \citealt{Majewski2014}), is one of the four Sloan Digital Sky Surveys III  (SDSSIII; \citealt{Eisenstein2011}).
Recently, \cite{Anders2014} have defined a subsample of the first year of APOGEE data (as part of data release 10, DR10; \citealt{Ahn2014}). The selection criteria for what the latter authors named their APOGEE High Quality 
Giant Sample are summarized in Table 1 of \cite{Anders2014}.    

\begin{figure}
\includegraphics[scale=.6,angle=0]{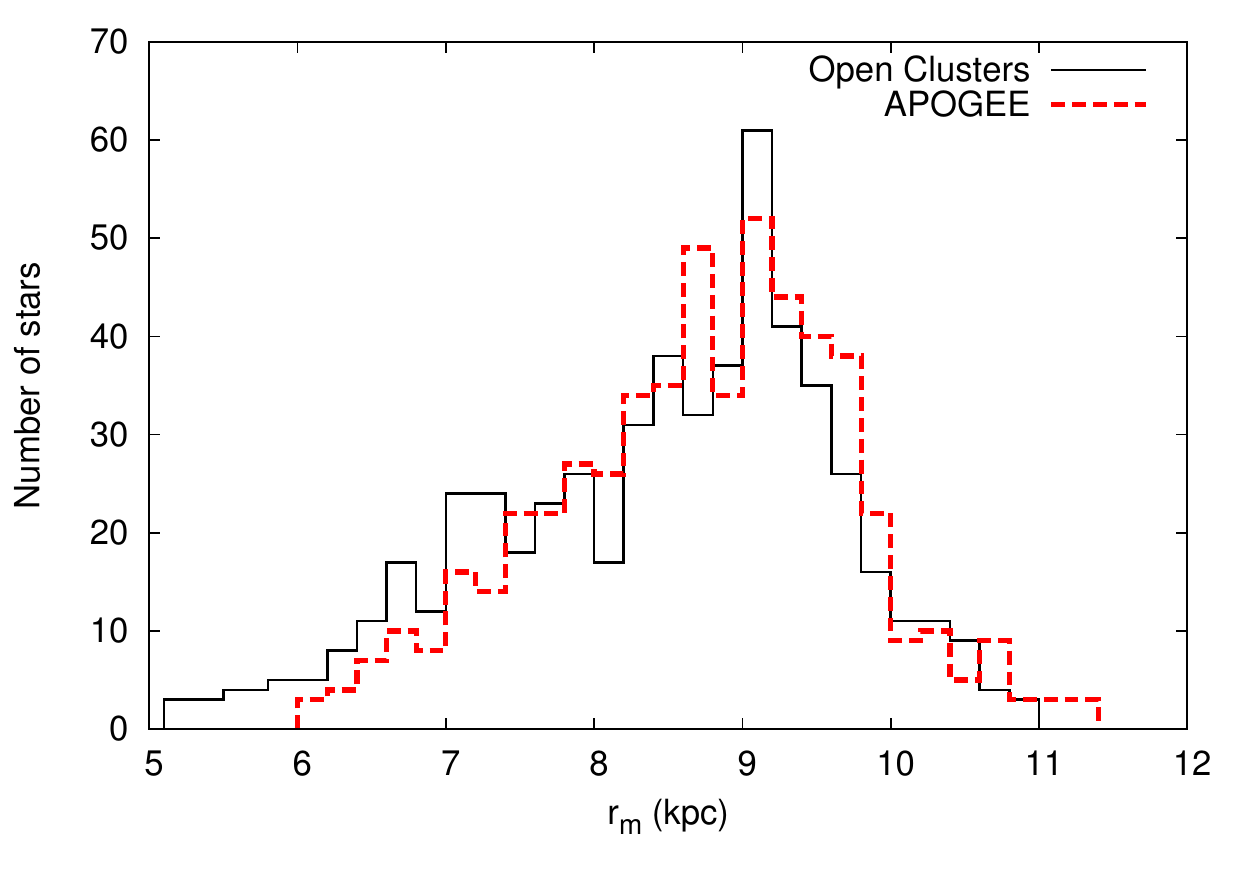}
\caption{This figure shows the mean radius distribution of the OC sample (solid black curve) and red giant branch stars from APOGEE (dashed red curve). Both samples have similar distribution, as well as almost the same 
number of objects per bin.}
\label{figX}
\end{figure}

Starting from a similar sample, we have selected stars that stay on the galactic plane (i.e. with their maximum vertical orbital amplitude, z$_{max}$, below 0.2 kpc. Moreover, we required a combined 
proper-motion error below 4 mas/yr ($\delta_{\mu_{RA}} < 4$ mas/yr and $\delta_{\mu_{DEC}} < 4$ mas/yr, mean proper motion error in right ascension and declination, respectively). The final sample resulted in 559 stars from DR10 which is a subsample from 
what \cite{Anders2014} named their gold sample. The distances were computed using the distance code of Santiago \citep[priv. communication - see also][]{Santiago2015}. The mean uncertainties of distances and proper 
motions for the APOGEE DR10 sample are shown by \cite{Anders2014}, where their gold sample have a threshold in uncertainties of 20$\%$ in distances and 4 mas/yr in proper motion. Also for the APOGEE sample, the uncertainties in 
radial velocities are less than $5\%$. Their mean radius distribution are shown in Fig.~\ref{figX} (dashed red histogram), and turned out to be similar to that of our sample of OCs. The main difference is that the APOGEE giants span, 
most probably a larger age range. Indeed, we expect stars with mean radius in the 7-10 kpc range to be predominantly of ages between 1 and 6 Gyrs \citep[see][]{Minchev2014}. For consistency, for the final sample of APOGEE stars 
we recompute the mean radius $r_m$ using the rotation curve given by Eq.~\ref{eq:v_L2011} instead of the one from \cite{Anders2014}. However, in both calculations the mean radii are very similar.  

\begin{figure}
\includegraphics[scale=.5,angle=0]{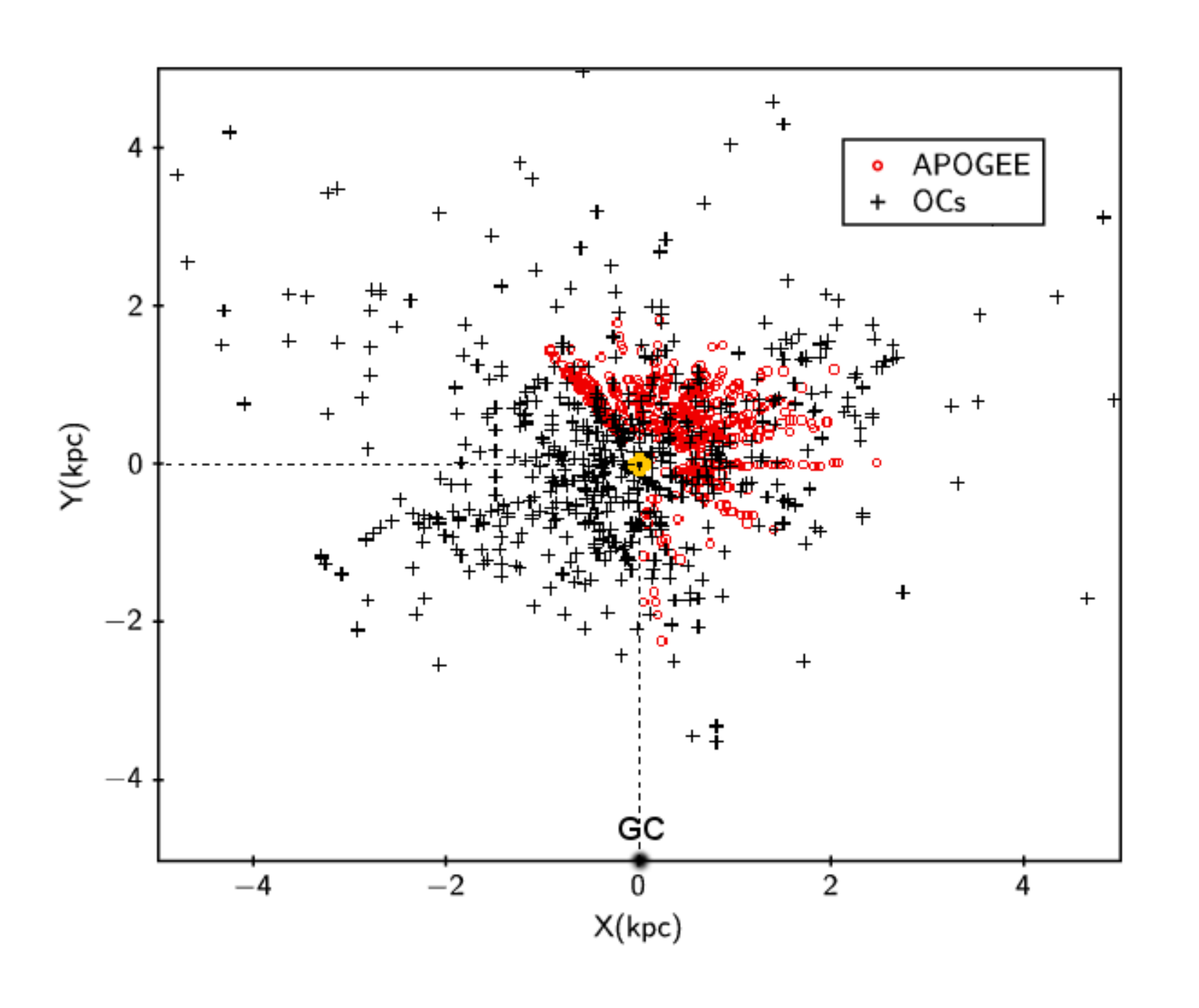}
\caption{This figure shows the distribution in the X-Y galactic plane, with the Sun (yellow dot) locate in (0,0), for our adopted OC sample (black crosses) and red giant branch stars from APOGEE (red open circles). The Galactic 
center (GC) is pointed toward negative values in the Y-direction indicated by a black dot. We can see that the OC sample has a more wide azimuthal distribution than the  giant branch stars from APOGEE (see text for more details).}
\label{figX2}
\end{figure}

Finally, in Fig.~\ref{figX2} we show the spatial distribution of both samples on the X-Y Galactic plane. The first thing that can be noticed is that the OCs are more homogeneously distributed in the azimuthal direction than the 
APOGEE sample adopted here. The second difference is that we have more OCs in the inner part of the disk, while the APOGEE sample is more concentrated in the outer parts. This will certainly be improved once APOGEE-2 data will be 
available. As we will see in the next Section, despite these main differences, the pattern speed computed with both samples turned out to be very similar.

\section{RESULTS AND DISCUSSION}
\label{res}

\begin{table*}
\caption{Compilation of some estimations of $\Omega_p$ found in the literature.}
\label{tb4}
\begin{center}
\begin{tabular}{|c|c|c|c|}
\toprule
$\Omega_p$ (km s$^{-1}$ kpc$ ^{-1})$ & Method* & Objects & Reference \\
\hline\hline
19.1$\pm$3.6 & 1 & Cepheids                        & \cite{Mishurov1979}\\
30$\pm$0.7   & 1 & O and B-type stars and Cepheid  & \cite{Fernandes2001}\\
20$\pm$2     & 2 & Open clusters              & \cite{Amaral1997}\\
24$\pm$1     & 2 & Open clusters              &\cite{Dias2005}\\
20.3$\pm$0.5 & 2 & Sample of runaways and early-type stars (from {\it Hipparcos})  &\cite{Silva2013}\\
21.2$\pm$1.1 & 2 & Only early-type stars      &\cite{Silva2013}\\
18.1$\pm$0.8 & 3 & {\it Hipparcos} subsample      &\cite{Quillen2005}\\
18.6$\pm$0.3 & 4 & RAVE survey  &\cite{Siebert2012}\\
\bottomrule
\end{tabular}
\end{center}
\begin{tabular}{|l|}
*method 1 = kinematic model, method 2 = Birthplace technique, method 3 = Orbital analysis of moving group in the (u-v) plane,\\
*method 4 = Spiral perturbation to reproduce the gradient in the mean galactocentric radial velocity.
\end{tabular}
\end{table*}

In Sec.~\ref{mtd} we described and tested a new method, without invoking any prior information about the spiral arms, which proved to be useful to constrain the value of the pattern speed within an error of $\sim 1$ km s$^{-1}$ kpc$^{-1}$. Here 
we show the results that we obtain by applying our new method to a sample of disk stars and open clusters, described in Sec.~\ref{data}.

\subsection{The value of \texorpdfstring{$\Omega_{p}$}{Op} and the corotation radius}

\begin{figure}
\includegraphics[scale=.6,angle=0]{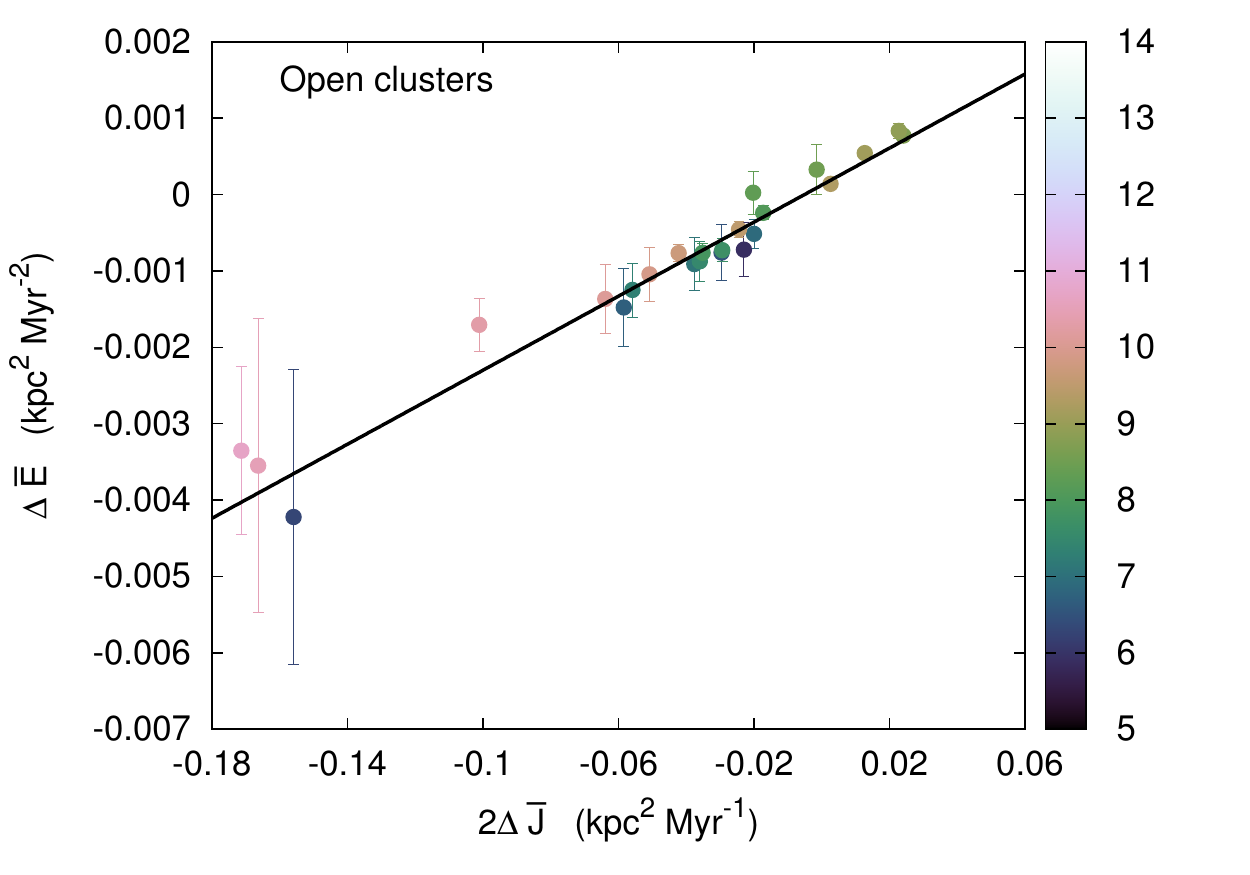}
\includegraphics[scale=.6,angle=0]{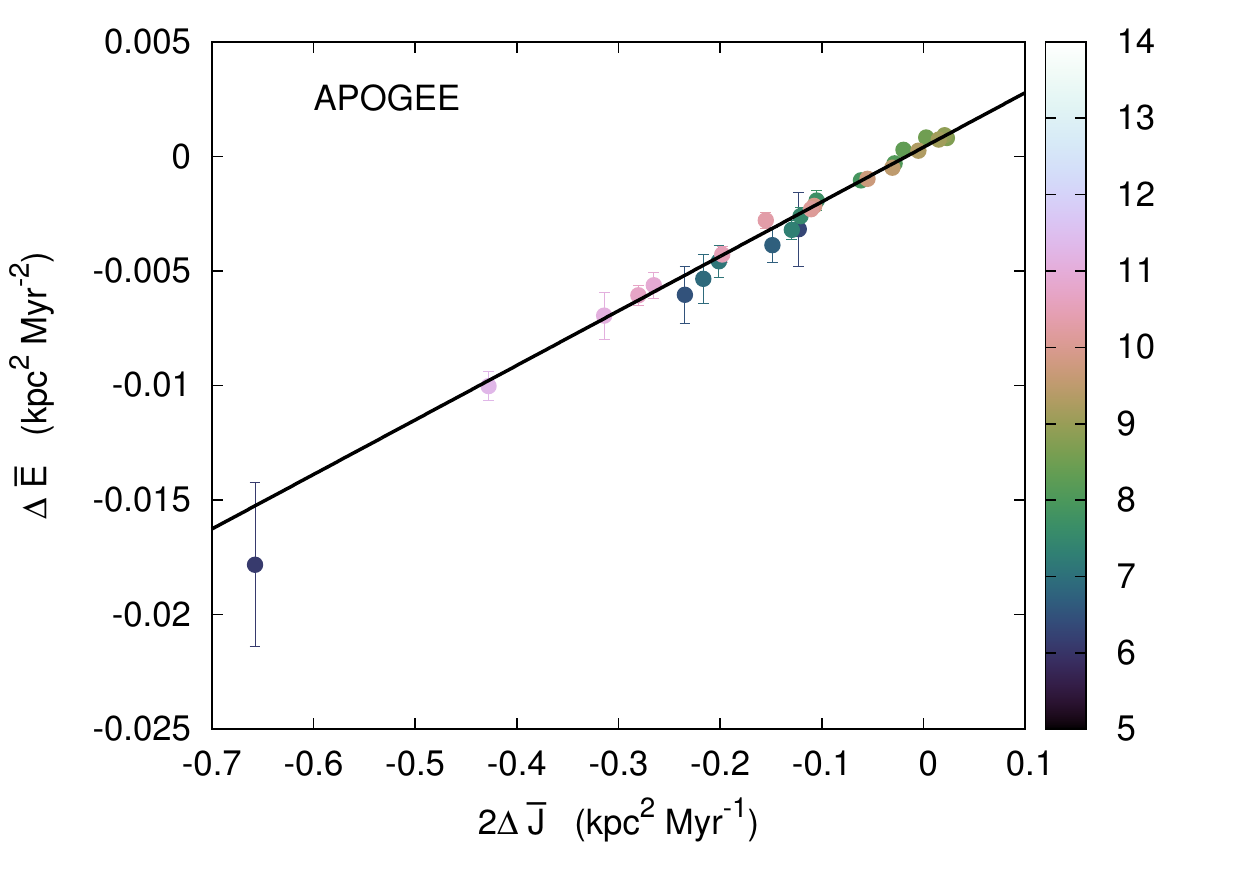}
\includegraphics[scale=.6,angle=0]{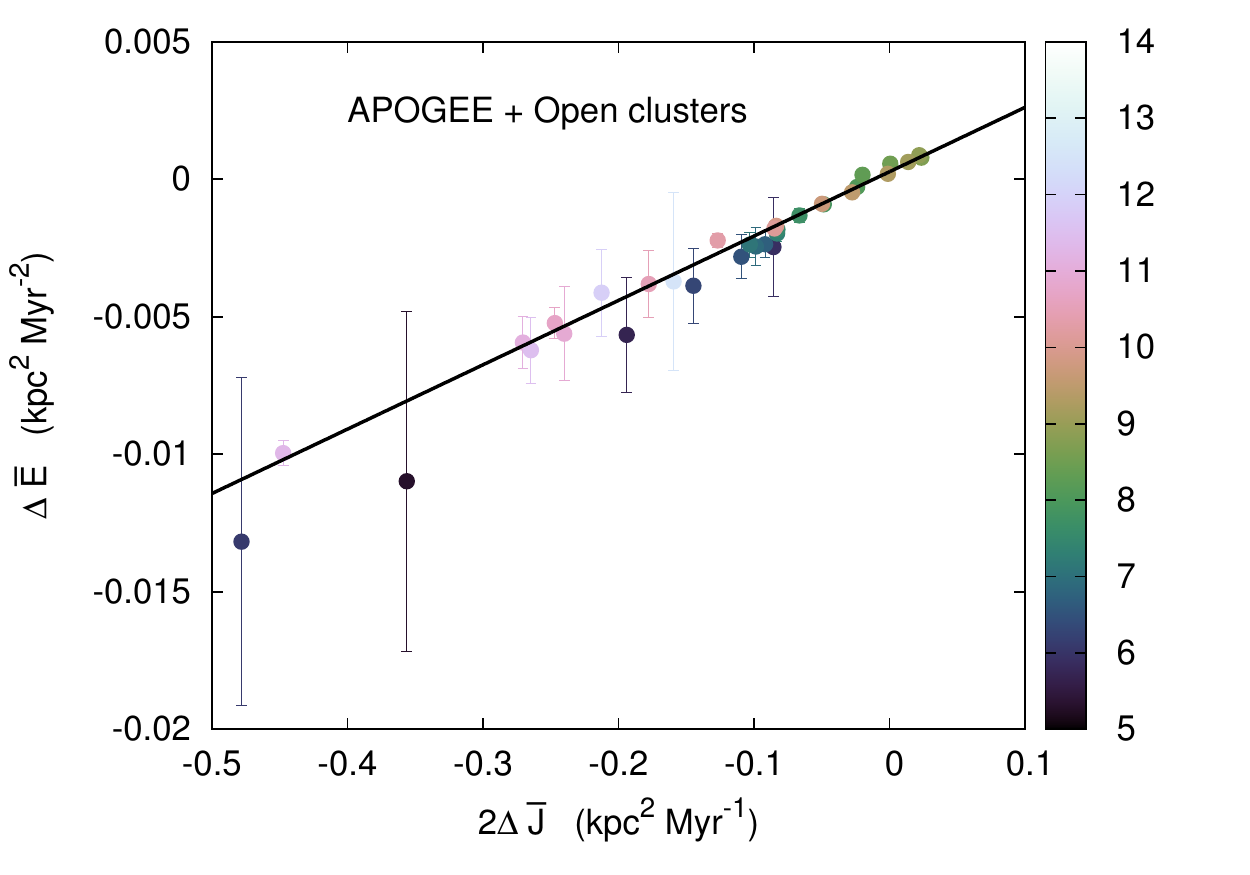}
\caption{This figure shows the linear correlation between $2\Delta \overline{J}$ and $\Delta \overline{E}$ for a real sample. The dots are the values computed from Eqs.~\ref{delem} and 
\ref{deljm}, as explained in Sec.~\ref{mtd}, and the color code represent the bin radius in which each dot was computed. The black lines are the least-squares fit on these points, where the slope gives us the value of $\Omega_p$. 
The upper panel shows the result for the OC sample; the middle shows the results for the APOGEE sample and the last one is the combination of both, APOGEE + OCs samples. The values of $\Omega_p$ found for each sample are shown on 
Tab.~\ref{tb3}.}
\label{fig3}
\end{figure}

Our main results are summarized in Fig.~\ref{fig3}, where we plot $2\Delta \overline{J}$ versus $\Delta \overline{E}$. The slope of the observed linear correlation gives the value of $\Omega_p$ (shown in Tab.~\ref{tb3}), 
as described by Eq.~\ref{op2}. The upper, middle and bottom panels show the results we obtain using the OC, red giants and the 
combination of the two, respectively. The color gradient indicates the center of the bin in radius in which each point. The value that we find by combining the two samples is 
$\Omega_p = 23.0\pm0.5$ km s$^{-1}$ kpc$^{-1}$. For $\Omega_p = 23$ km s$^{-1}$ kpc$^{-1}$ the corotation radius in our model is situated at r$_{cr} = (1.09\pm0.04)$R$_0$ which is in agreement with results found by \cite{Dias2005} 
with r$_{cr} = (1.06\pm0.08)$R$_0$.  The fact that the Galaxy presents a well defined corotation radius supports the idea of a dominant pattern speed, at least, for an old stellar population (see further discussion in Section 4.3). 
It's interesting to notice that the values obtained for $\Omega_p$ from the fit, in Fig.~\ref{fig2} or Fig.~\ref{fig3}, have unit of Myrs$^{-1}$. However the usual unit given for the pattern speed is in km s$^{-1}$ kpc$^{-1}$, thus 
to transform from (Myrs$^{-1}$) to (km s$^{-1}$ kpc$^{-1}$) we must divide by a factor of 0.00102.

We should keep in mind that our quoted uncertainties take into account only the fitting procedure. Systematic and intrinsic errors (e.g. assumptions made on 
the method and variation of the rotation curve because of uncertainties on the solar parameters $R_0$ and $V_0$) were not taken into account. However, we can estimate that by varying the values of $R_0$ and $V_{0}$ between $7-8.5$ 
kpc and $199-245$ km s$^{-1}$ respectively. These variations would propagate into a $\sim$2 km s$^{-1}$ kpc$^{-1}$ change on the final value of $\Omega_p$. \cite{Piffl2014} recently found $V_{0} = 240$ km s$^{-1}$ for $R_0 = 8.3$ kpc, 
which is compatible with our adopted values and hence will change our estimate of $\Omega_p$ within the intrinsic errors. In Sec.~\ref{Am} we discussed more about the intrinsic uncertainties of our method.

As we can see in the last panel in Fig.~\ref{fig3} the combination of both data, APOGEE $+$ OCs, has a better distribution in $2\Delta \overline{J}$, this is due to a better spatial distribution when we combine 
the data. It happens because the stars have different values of radial velocities (and they do not belong to a group in a U-V velocity diagram) given us a better average of the angular momentum variation in an annular region of the disk. Thus, we expected that with the new coming data, from APOGEE-2 
(now starting with SDSSIV), our results will improve, since it  will increase the number of stars and in a way that the distribution in azimuth will become more homogenous. Another thing we can notice in this figure is that the 
points with higher  $|2\Delta \overline{J}|$ and $|\Delta \overline{E}|$ come from mean radius around 6 or 12 kpc, that are very close to the IRL and OLR resonances and as we discussed before this regions warm up the disk. 

\subsection{Comparison with previous reported results in the literature}

One of the most important parameters in studying the spiral structure is its pattern speed $\Omega_p$. Although the fundamental nature of the spiral arms is not fully understood it plays an important role in galactic dynamics 
\citep[e.g.][]{Antoja2009,Lepine2011b,Quillen2011,Minchev2012,Sellwood2014} and its pattern speed is a fundamental parameter that drives all the resonances in the disc. Table~\ref{tb4} summarizes the several previous attempts 
made in the literature to estimate the value of $\Omega_p$. \cite{Gerhard2011} made a review from the values found for the pattern speeds in the Milky Way and he end up with a range between $\Omega_p\sim17-30$ km s$^{-1}$ kpc$^{-1}$.

As it is clearly seen from Table~\ref{tb4} and \ref{tb3}, our results are in agreement with studies that suggest a pattern speed between 20-25 km s$^{-1}$ kpc$^{-1}$. However, \cite{Quillen2005} and \cite{Siebert2012} found values bellow 
20 km s$^{-1}$ kpc$^{-1}$, which are not in agreement with our results even with error bars around $\sim$ 3 km s$^{-1}$ kpc$^{-1}$. In the literature we see that higher values for $\Omega_p$ are preferred by open cluster birthplaces 
while hydrodynamical simulations and phase space substructures favor slower pattern speeds. Thus, since \cite{Quillen2005} and \cite{Siebert2012} analysis are based on moving groups and the velocity gradient both close to the solar 
neighborhood, which are substructures and can be associated only with one spiral structure (as e.g., Perseus arm or even a local spur arm), it could explain why they found lower values for $\Omega_p$. The discrepancies of values 
found in the literature are discussed in Sec.~\ref{mps}, as a possible contamination by multiple spiral patterns, which are difficult to be taken into account and can lead to a systematic errors that explain the wide range of values 
found for $\Omega_p$, depending on the tracers and the methods that were used to estimate it. 

The value that we found for the dominant MW spiral pattern speed is also in agreement with the values found in many external galaxies. For example, \cite{Scarano2013} found $\Omega_p$ to have a distribution concentrated 
around 24 km s$^{-1}$ kpc$^{-1}$. However, the fundamental nature of the pattern speed is still not clear, which requires more theoretical work to be fully understood.  

\begin{table}
\caption{This table contains our results for the pattern speed $\Omega_p$ obtained from two different samples and the combination of all them.}
\label{tb3}
\begin{center}
\begin{tabular}{*3c}
\toprule
&$\Omega_p$(km s$^{-1}$ kpc$ ^{-1}$) & Sample \\
\hline\hline
&24.0$\pm$1.0 & Open clusters\\
&23.3$\pm$0.6 & Apogee\\
&23.0$\pm$0.5 & Apogee + Open clusters\\
\bottomrule
\end{tabular}
\end{center}
\end{table}

\subsection{Possible contamination by multiple spiral patterns}
\label{mps}

\cite{Valle2014} uses many tracers to probe the spiral structure of our Galaxy and concludes that it has a four-armed spiral pattern. However, only two of these may be present in the density distribution of old stars 
\citep[see, e.g.][]{Drimmel2000,Martos2004}. Multiple spiral patterns could possibly be a source of errors when we try to determine only one pattern speed. \cite{Naoz2007} measured the pattern speed for: Sagittarius-Carina and 
they found a superposition of two pattern speeds with $\Omega_p = 16.5\pm2.0$ km s$^{-1}$ kpc$^{-1}$ and $\Omega_p = 29.8\pm2.0$ km s$^{-1}$ kpc$^{-1}$, Perseus arm $\Omega_p = 20.0\pm2.0$ km s$^{-1}$ kpc$^{-1}$ and 
Orion $\Omega_p = 28.9\pm2.0$ km s$^{-1}$ kpc$^{-1}$. Some models also support multiple pattern speeds in order to explain radial migration in discs of galaxies \citep{Minchev2006,Minchev2012,Grand2014}. However others 
studies suggest that the MW has a corotation radius well established, situated close to the solar radius \citep[][among others]{Marochnik1972,Creze1973,Mishurov1999,Dias2005,Amores2009}. It would be possible only if we have a dominant 
pattern speed rotating rigidly, nevertheless this do not exclude small structures to exist, that may rotate with different angular velocities. For $\Omega_p = 23$ km s$^{-1}$ kpc$^{-1}$ the corotation radius in our model is situated at 
r$_{cr} = (1.09\pm0.04)$R$_0$ which is in agreement with results found by \cite{Dias2005} with r$_{cr} = (1.06\pm0.08)$R$_0$.  The fact that the Galaxy presents a well defined corotation radius supports the idea of a dominant 
pattern speed, at least, for an old stellar population. Therefore, one way to avoid a possible contamination by different pattern speeds is to split the sample into old and young stars.  

In the future we will apply our method to a N-body simulation sample with multiple pattern speeds to check if we are able to distinguish multiple arms and/or analyze the influence of small arms on the dominant ones.

\section{Conclusions}

In this work, we proposed a new method to derive the spiral pattern speed of the MW based on the interaction between the spiral arms and the stellar objects. In this method we do not need any prior information about the spiral arms, 
as for example its shape and location. In addition, we do not need to be concerned with the stellar ages, which allows the use of data from large Galactic surveys. 

The assumption we make in our method is that the initial energy and angular momentum of the objects can be approximated as the circular orbit at the mean radius. This approach introduces a natural error on the order of 
1$\sim$2 km s$^{-1}$ kpc$^{-1}$ which is equivalent, or even smaller, than the available methods to constrain the value of $\Omega_p$.

Using a sample of open clusters and red giant stars from the APOGEE DR10 \citep{Ahn2014} we have found $\Omega_p = 23.0\pm0.5$ km s$^{-1}$ kpc$^{-1}$, which is compatible with other values in literature and placed the corotation radius at 
r$_{cr} = 8.74$ kpc for a solar position R$_0 = 8$ kpc. We have to stress again that the given error for $\Omega_p$ here is just due to the RMS from the fitting procedure. A more realistic error estimate which also takes into 
account the errors intrinsic to our method should  be around 2 km s$^{-1}$ kpc$^{-1}$ (intrinsic method error + RMS). Systematic errors and errors due to other sources of perturbation (as discussed in Sec.~\ref{mps}) 
are even more difficult to estimate, and could most probably explain the range of values for $\Omega_p$, between $17-30$ km s$^{-1}$ kpc$^{-1}$, found in the literature. Further studies, using N-body simulations data, 
are needed to check the effective influence of multiple spiral arms on the determination of $\Omega_p$, assuming a constant pattern speed.    

The new method for estimating the spiral pattern speed presented here can be tested with the large amounts of currently available data of ever increasing quality from large Galactic surveys, such as RAVE \citep{Steinmetz2006}, 
SEGUE \citep{Yanny2009}, APOGEE \citep{Allende2008,Majewski2014}, GES \citep{Gilmore2012}, and in the near future - Gaia \citep{Bruijne2012}, 4MOST \citep{Jong2012} and WEAVE\citep{Dalton2012}.

\section*{Acknowledgments}

I would like to tanks Douglas A. Barros for the comments and helpful discussions to improve this paper.  
TCJ is supported by DAAD-CNPq-Brazil through a
fellowship within the program "Science without Borders".


\bsp

\label{lastpage}

\end{document}